\documentclass[12pt,reqno]{article}
\pdfoutput=1
\usepackage{amsmath,hyperref,amssymb}
\usepackage{subfigure,graphicx,url}
\usepackage{slashed}

\usepackage{tabularx}
\usepackage{bm}
\usepackage{euscript}
\usepackage{graphicx}
\usepackage{color}
\usepackage{amsfonts}
\usepackage{exscale}
\usepackage{amsbsy}
\usepackage{subfigure}
\usepackage{textcomp}

\newcommand{\be}{\begin{eqnarray}}
\newcommand{\ee}{\end{eqnarray}}

\newcommand{\bea}{\begin{eqnarray}}
\newcommand{\eea}{\end{eqnarray}}
\newcommand{\eeq}{\end{equation}}
\newcommand{\beq}{\begin{equation}}

\newcommand{\mc}{\mathcal}
\renewcommand{\t}{\tilde}

\numberwithin{equation}{section}

\DeclareSymbolFont{AMSa}{U}{msa}{m}{n}
\DeclareSymbolFont{AMSb}{U}{msb}{m}{n}
\DeclareMathSymbol{\fieldR}{\mathalpha}{AMSb}{"52}

\def\beq{\begin{equation}}
\def\eeq{\end{equation}}
\def\be{\begin{equation}}
\def\ee{\end{equation}}
\def\bea{\begin{eqnarray}}
\def\eea{\end{eqnarray}}

\makeatletter
\renewcommand\section{\@startsection {section}{1}{\z@}%
                                 {-3.5ex \@plus -1ex \@minus -.2ex}
                                   {2.3ex \@plus.2ex}%
                                   {\normalfont\large\bfseries}}
\renewcommand\subsection{\@startsection{subsection}{2}{\z@}%
                                   {-3.25ex\@plus -1ex \@minus -.2ex}%
                                     {1.5ex \@plus .2ex}%
                                     {\normalfont\bfseries}}
\renewcommand\subsubsection{\@startsection{subsubsection}{3}{\z@}%
                                   {-3.25ex\@plus -1ex \@minus -.2ex}%
                                     {1.5ex \@plus .2ex}%
                                     {\normalfont\itshape}}
\makeatother

\begin{document}

\setcounter{page}1

\begin{flushright} \small
SU-ITP-13/15
\end{flushright}
\bigskip

\begin{center}
 {\LARGE\bfseries Supersymmetric Defect Models and \\[3mm] Mirror Symmetry }\\[3mm]

 \

Anson Hook$^\dag$, Shamit Kachru$^*$ and Gonzalo Torroba$^{*}$   \\[5mm]
 
 {\small\slshape
$^\dag$ School of Natural Sciences\\
 Institute for Advanced Study\\
Princeton, NJ 08540, USA\\
\medskip
 $^*$ Stanford Institute for Theoretical Physics,\\
 Department of Physics\\
 and\\
 Theory Group, SLAC \\
 Stanford University\\
 Stanford, CA 94305, USA \\
\medskip
 {\upshape\ttfamily hook@ias.edu, skachru@stanford.edu, torrobag@stanford.edu}
\\[3mm]}
\end{center}

\

\vspace{1mm}  \centerline{\bfseries Abstract}
\medskip
We study supersymmetric field theories in three space-time dimensions doped by various configurations of electric charges or magnetic fluxes.  These are supersymmetric avatars of impurity models.  In the presence of additional sources such configurations are shown to preserve half of the supersymmetries.
Mirror symmetry relates the two sets of configurations.  We discuss the implications for impurity models in 3d ${\cal N}=4$ QED with a single charged hypermultiplet (and its mirror, the theory of a free hypermultiplet) as well as 3d ${\cal N}=2$ QED with one flavor and its dual, a supersymmetric Wilson-Fisher fixed point.  Mirror symmetry allows us to find backreacted solutions for arbitrary arrays of defects in the IR limit of ${\cal N}=4$ QED.  Our analysis, complemented with appropriate string theory brane constructions, sheds light on various aspects of mirror symmetry, the map between particles and vortices and the emergence of ground state entropy
in QED at finite density.

\thispagestyle{empty}

\bigskip
\newpage

\addtocontents{toc}{\protect\setcounter{tocdepth}{2}}

\tableofcontents

\vskip 1cm

\section{Introduction}\label{sec:intro}

Duality is a powerful tool in analyzing quantum field theories.  An early and surprising manifestation was the discovery of the relationship between the XY model and the abelian Higgs model in 
2+1 dimensions \cite{Peskin,Dasgupta}.  This duality has been generalized to models with 
additional `flavors' of matter fields in \cite{Motrunich,Balents}, and plays a role in the study of 
lattice models of antiferromagnets.  

The study of these field theories in the presence of external charges (e.g., impurities) is of considerable interest.   A single defect interacting with a Wilson-Fisher critical theory was discussed in \cite{Buragohain}.  In general it is a tough question to determine the backreaction of a given configuration of defects on the bulk field theory, and even tougher to compute quantities of interest for IR transport, like current correlators in a lattice of defects.

Here, we focus on supersymmetric field theories.  Theories with 3d ${\cal N}=4$ and ${\cal N}=2$ supersymmetry were
shown to enjoy a mirror symmetry~\cite{Mirror, deBoer:1996mp,deBoer:1996ck,deBoer:1997kr,Aharony:1997bx,Kapustin:1999ha}, where two distinct UV gauge theories flow to the same IR fixed point. Mirror symmetry is in many ways a supersymmetric cousin of the XY / abelian Higgs duality \cite{Yin}, and so many of the questions of
interest in that context can be imported to 3d supersymmetric mirror pairs.  The additional theoretical tools afforded by supersymmetry allow some of these questions to be answered.

In this paper, we focus on questions of charged defect or impurity physics in these supersymmetric theories; earlier work in this spirit, where the focus was on holographic supersymmetric constructions, includes \cite{Karch,Jensen,Harrison,Benincasa}.
We study both electric and magnetic impurities in the simplest mirror pairs (reviewed in \S \ref{sec:susy-review}) of theories
with ${\cal N}=4$ and ${\cal N}=2$ supersymmetry.
We will see in \S\S \ref{sec:defectsN4} and \ref{sec:defectsN2} that such impurities can preserve 1/2 of the supersymmetry in the presence of appropriate external backgrounds for additional fields.  In particular, we will show that it is possible to preserve supersymmetry at finite density for local and global $U(1)$ symmetries.

We then find that the power of mirror symmetry allows us to extract non-trivial information about the IR nature of the solution for bulk fields in the
presence of defects, and even (in the original ${\cal N}=4$ example) allows us to compute in \S \ref{sec:lessons} the lattice backreaction at strong coupling. This gives a promising method for studying lattices of impurities interacting with strongly coupled field theories.  Besides applications to such systems, our results contribute to the understanding of various formal aspects of mirror symmetry, and provide a more explicit map between particles and vortices.

At some points it is useful to make contact with string theory.  Various questions that have arisen in studies of holography (such as the finite ground-state entropy of certain doped field theories) can be viewed in a different light in our constructions, along the lines
envisioned in \cite{Kraus,Polchinski}. In \S \ref{sec:nonsusy} we show that our construction explains  the emergence of ground-state degeneracy in strongly interacting QFTs at finite density. Specifically, the $\mc N=4$ SQED theory at finite chemical potential for the topological $U(1)$ symmetry (defined in (\ref{eq:topologicalJ}) below) is equivalent in the IR to free electrons in an external magnetic field. The ground state entropy comes from the Landau level degeneracy of the dual.
On the other hand, the construction
of 3d supersymmetric gauge theories via brane configurations (following \cite{HananyWitten})
makes manifest many of the properties of mirror symmetry in the presence of defects.  This is studied in \S \ref{sec:Dbrane}. Finally, \S \ref{sec:discussion} suggests various future directions motivated by our results.

\section{Three dimensional theories and mirror symmetry}\label{sec:susy-review}

Here, we discuss the field content and Lagrangians of the theories we'll be interested in throughout the rest of the paper. These are three-dimensional field theories with $\mc N=2$ and $\mc N=4$ supersymmetry, namely 4 and 8 supercharges respectively. We will do this in terms of ${\cal N}=2$ multiplets, since these follow from dimensional reduction of the familiar
 4d ${\cal N}=1$ multiplets. Theories with $\mc N=2$ have simpler matter content than their $\mc N=4$ counterparts, but their infrared dynamics is richer and more involved. For this reason, we will first consider $\mc N=4$ theories. We follow the original works on the subject~\cite{Mirror, deBoer:1996mp,deBoer:1996ck,deBoer:1997kr,Aharony:1997bx,Kapustin:1999ha}.
 
\subsection{Three-dimensional supersymmetric theories}\label{subsec:3dsusy}

To begin, we review the field content of the 3d ${\cal N}=4$ multiplets. 
An ${\cal N}=4$ hypermultiplet ${\cal Q}$ consists of a pair of ${\cal N}=2$ chiral superfields, $Q$ and $\tilde Q$, in conjugate
representations of the gauge group.  The ${\cal N}=4$ vector multiplet ${\cal V}$ consists of an 
${\cal N}=2$ vector superfield $V$ and an ${\cal N}=2$ chiral superfield $\Phi$.  Recall as well that
the ${\cal N}=2$ vector superfield contains a real scalar
field $\sigma$ (the extra gauge field component in the dimensional reduction from 4d), so each ${\cal N}=4$ vector multiplet gives rise to a triplet of scalar fields.

Our notation for the component fields in general will be as follows.  The Bose components of $V$ consist of the gauge field $A_{\mu}$, the scalar $\sigma$ mentioned above, and an auxiliary field $D$.
The gaugino $\lambda$ is its Fermi component.  $\Phi$ contains a complex scalar $\phi$ as well as fermions $\psi$ and an auxiliary field $F$.  The scalars in $Q, \tilde Q$ will be denoted by 
$q, \tilde q$, while the fermions will be $\psi_q, \psi_{\t q}$ and so forth.  We work in $(+--)$ signature and follow the conventions for three dimensional SUSY of~\cite{Schwarz:2004yj}.

The simplest $\mc N=4$ theory is that of a free hypermultiplet $\mc Q = (Q, \t Q)$. In $\mc N=2$ notation,
\be
L_H(\mc Q)= \int d^4 \theta (Q^\dag Q + \t Q^\dag \t Q)= |\partial_\mu q|^2 + |\partial_\mu \t q|^2+ i \bar \psi_q \gamma^\mu \partial_\mu \psi_q+i \bar \psi_{\t q} \gamma^\mu \partial_\mu \psi_{\t q}\,.
\ee
We always take the fermions to be Dirac, and $\bar \psi = \psi^\dag \gamma^0$.
This theory has a global $U(1)$ symmetry under which $Q$ has charge $+1$ and $\t Q$ has charge $-1$. Since we will be interested in understanding the effects of $x$-dependent background fields (related to the insertion of defects), let us introduce a background vector multiplet $\hat{ \mc  V}=(\hat V, \hat \Phi)$ for the $U(1)$ symmetry of this simple theory. This modifies the Lagrangian to
\be\label{eq:LQ}
L_H(\mc Q, \hat{\mc V})= \int d^4 \theta (Q^\dag  e^{2\hat V}Q + \t Q^\dag e^{-2 \hat V} \t Q) + \int d^2 \theta \, \sqrt{2} \hat \Phi Q \t Q + c.c.
\ee
In components,
\bea
L_H(\mc Q ,\hat{\mc V})&=& |\hat D_\mu q|^2 + |\hat D_\mu \t q|^2+ i \bar \psi_q \not\!\!\hat D \psi_q+i \bar \psi_{\t q} \not \!\! \hat D \psi_{\t q}- (\hat \sigma^2+ 2|\hat \phi|^2) (|q|^2+ |\t q|^2)   \nonumber\\
&-&\hat \sigma (\bar \psi_q \psi_q - \bar \psi_{\t q} \psi_{\t q})   -\sqrt{2}(\hat \phi \psi_q \psi_{\t q}+ c.c.)+ \sqrt{2}\left(i \bar{\hat \lambda}(q^\dag \psi- \t q^\dag \psi_{\t q})+c.c.\right) \nonumber\\
&+&\hat D (|q|^2 - |\t q|^2)+ \sqrt{2} (\hat F q \t q + c.c.)
\eea
where the covariant derivative
\be
\hat D_\mu \varphi_i \equiv (\partial_\mu + i e_i \hat A_\mu) \varphi_i
\ee
for a field $\varphi_i$ of charge $e_i$. The background scalars $\hat \sigma$ and $\hat \phi$ give real and complex masses, respectively; we also note the possibility of background D- and F-terms, that will appear in our analysis in later sections.

Next, let us consider an $\mc N=4$ $U(1)$ gauge theory with vector-multiplet $\mc V$ and a charged hypermultiplet $\mc Q$. The Lagrangian is
\be
L = L_H(\mc Q, \mc V) + L_V(\mc V)
\ee
where $L_H(\mc Q, \mc V)$ is given by (\ref{eq:LQ}) with the replacement $\hat{\mc V} \to \mc V$, and the kinetic terms for the vector superfield are
\bea
L_V(\mc V) &=& \frac{1}{2g^2}\int d^2\theta \,W_\alpha W^\alpha + \frac{1}{g^2}  \int d^4 \theta \,\Phi^\dag \Phi \\
&=& \frac{1}{g^2} \left[-\frac{1}{4} F_{\mu\nu}^2+ \frac{1}{2} (\partial_\mu \sigma)^2 +|\partial_\mu \phi|^2+ i \bar \lambda \not\! \partial \lambda+  i \bar \psi_\phi \not\! \partial \psi_\phi+\frac{1}{2} D^2+|F|^2 \right] \nonumber\,,
\eea
where $F_{\mu\nu}=\partial_\mu A_\nu - \partial_\nu A_\mu$.
The Lagrangian for more general ${\cal N}=4$ field contents can be found in the obvious way.
For instance, in the event that there are multiple hypermultiplet flavors one simply adds an $i$ index to the
$q$ and $\tilde q$ fields with $i=1, \cdots, N_f$.

There is one further symmetry that will be useful to keep in mind.  In three dimensions, any abelian gauge field gives rise to a global $U(1)_J$ current via the relation
\begin{equation}\label{eq:topologicalJ}
J_{\mu} =\frac{1}{2} \epsilon_{\mu\nu\rho}F^{\nu\rho}~.
\end{equation}
It is clear that the `charge' $J_0$ of this $U(1)_J$ symmetry is carried by configurations with nonzero
magnetic flux -- i.e., vortices.
This symmetry shifts the dual photon
\be\label{eq:dualphoton}
F_{\mu\nu} =\epsilon_{\mu\nu\rho}\partial^\rho \gamma
\ee
by a constant.\footnote{With this normalization, the dual photon has kinetic term $L \supset \frac{1}{2g^2}(\partial_\mu \gamma)^2$. Also, charge quantization (see e.g.~\cite{Seiberg:1996nz}) implies that it is compact with period
$\gamma \to \gamma +g^2$. } This symmetry plays a central role in mirror symmetry, as we review shortly. We can consider turning on a background vector superfield $\hat {\mc V}$ for the global $U(1)_J$, which then couples to $\mc V$ via a BF interaction,
\be\label{eq:BF}
L_{BF}(\mc V, \hat{\mc V}) =\frac{1}{2\pi}\left( \frac{1}{2} \epsilon^{\mu\nu\rho} \hat A_\mu  F_{\nu\rho}+\sigma \hat D + \hat \sigma D+ \phi \hat F + \hat \phi F + \text{fermions} + c.c.\right)\,.
\ee
The background $ \hat A_\mu$ gives a chemical potential or magnetic flux source for the dynamical gauge field, and $\hat \sigma$ is an FI term. The other contributions are supersymmetric generalizations of these. Note that $L_{BF}$ is the supersymmetric version of coupling the dual photon to an external gauge field, $L \supset J_\mu \hat A^\mu$, with $J_\mu$ defined in (\ref{eq:topologicalJ}).

The $\mc N=2$ gauge theory can be obtained from the $\mc N=4$ version by erasing the chiral multiplet $\Phi$. It is also useful to connect the two theories by RG flows. Starting from the $\mc N=4$ theory, we can add a chiral superfield $S$ and couple it supersymmetrically to $\phi$ via the superpotential $W = S \Phi$. This makes $S$ and $\Phi$ massive, and in the IR we obtain the $\mc N=2$ gauge theory. Similarly, starting from the $\mc N=2$ theory we can add $\Phi$ and perturb by the superpotential $W =  Q\Phi \t Q$; this interaction is relevant and makes the theory flow to a point with enhanced $\mc N=4$ supersymmetry.

Finally, since we will discuss theories at finite density and/or magnetic field that respect some supersymmetry, we will need the supersymmetry variations. The variations for an $\mc N=2$ supersymmetry (Dirac) spinor $\epsilon$ are as follows. For a chiral superfield $\Phi = ( \phi, \psi, F)$,
\bea\label{eq:susychiral}
\delta \phi&=& \bar \epsilon \psi \nonumber\\
\delta \psi &=& (-i \not \! \! D \phi - \sigma \phi) \epsilon + F \epsilon^\dag \\
\delta F &=& \bar \epsilon^\dag (-i \not \! \! D \psi + \sigma \psi+ i \lambda \phi )\nonumber\,,
\eea
where $D_\mu$ is the covariant derivative introduced before. For a vector superfield $V= (A_\mu,\sigma, \lambda, D)$, the variations are
\bea\label{eq:susyvector}
\delta A_\mu &=& \frac{i}{2} \left( \bar \epsilon \gamma_\mu \lambda - \bar \lambda \gamma_\mu \epsilon\right) \nonumber\\
\delta \sigma&=& \frac{i}{2} \left( \bar \epsilon  \lambda - \bar \lambda  \epsilon\right) \nonumber\\
\delta \lambda &=& \left(\frac{1}{2} \gamma^\mu \gamma^\nu F_{\mu\nu}- \not \! \partial \sigma- i D \right)\epsilon \\
\delta D&=& \frac{1}{2} (\bar \epsilon \not \! \partial \lambda + \partial_\mu \bar \lambda \gamma^\mu \epsilon)\nonumber\,.
\eea

\subsection{Mirror symmetry for ${\cal N}=4$ theories}\label{subsec:mirrorN4review}

The mirror pair of theories we will be interested in is the pair given by the ${\cal N}=4$ abelian gauge theory with
one charged hyper ${\cal Q}$ on the one hand, and the free theory of an ${\cal N}=4$ hyper (which we denote by $\t{\mc Q}=(V_+, V_-)$) on
the other.  We will refer to these as the ``electric'' and ``magnetic'' theories, respectively.
This is the simplest example of~\cite{Mirror}, and mirror symmetry relating these theories
can be proven along the lines of~\cite{Kapustin:1999ha}.  Mirror symmetry states that both theories flow to the same infrared fixed point, so that their partition functions with external sources become equal:
\be\label{eq:mirror-map1}
Z_\text{electric}[\hat{\mc V}] = Z_\text{magnetic}[\hat {\mc V}]\,.
\ee
Here
\be
Z_\text{electric}[\hat {\mc V}] = \int D\mc V\,D \mc Q\, e^{i\int d^3x\,\left(L_H(\mc Q, \mc V)+ L_V(\mc V) + L_{BF}(\mc V, \hat {\mc V})\right)}
\ee
and
\be
 Z_\text{magnetic}[\hat {\mc V}]=\int D\t{\mc Q}\,e^{i\int d^3x\,L_H(\t{\mc Q}, \hat {\mc V})}\,.
\ee

Let us discuss the implications of the duality in more detail. ${\cal N}=4$ theories enjoy a global $SU(2)_L \times SU(2)_R$ R-symmetry.  In the abelian gauge theory, $SU(2)_L$ acts on the three scalars in the ${\cal N}=4$ gauge multiplet as a triplet, while leaving $q, \tilde q$ invariant; $SU(2)_R$ acts on $(q, \tilde q^*)$ as a doublet.  In the free hypermultiplet theory, $(v_+, v^*_-)$ form a doublet of $SU(2)_R$. Mirror symmetry exchanges $SU(2)_L$ in the QED theory with $SU(2)_R$ in the free hypermultiplet theory. It also maps the external sources $\hat{\mc V}$ according to (\ref{eq:mirror-map1}). In particular, the electric theory has a triplet of FI terms $(\hat \sigma, \hat \phi)$, which are spurions transforming under $SU(2)_R$; they are mapped to the real and complex masses in the magnetic theory, which are spurions of the $SU(2)_L$. In the mirror description, $U(1)_J$ acts simply as $\pm 1$ on the $V_{\pm}$ chiral
multiplets.  In this sense, mirror symmetry acts as particle/vortex duality.

In order to see how the moduli spaces map, recall that the gauge theory in the pair enjoys a scalar potential
\begin{equation}
\label{nfourscalarpot}
V = \frac{1}{2}g^2\left(|q|^2 - |\tilde q|^2+ \frac{\hat \sigma}{2\pi}\right)^2+g^2 \Big|\sqrt{2} q \t q+ \frac{\hat \phi}{2\pi} \Big|^2 +(\sigma^2+ 2| \phi|^2) (|q|^2+ |\t q|^2) \,.
\end{equation}
The first two terms come from integrating out the auxiliary fields in the vector multiplet, while the last terms come from the F-terms of the matter fields and the interaction with $\sigma$ can be understood as coming from the fourth component of the gauge field in 4d. In $\mc N=2$ language, the F-terms arise from the superpotential $W = \sqrt{2}\,\tilde Q \Phi Q$.
The effect of the background FI terms $(\hat \sigma, \hat \phi)$ is included for later applications. The theory has a Coulomb branch of vacua parametrized by $\phi$, $\sigma$ and the dual photon $\gamma$ defined in (\ref{eq:dualphoton}).  There is no Higgs branch with $N_f=1$ hypermultiplet, the case we are focusing on. In addition, quantum-mechanically the origin of the Coulomb branch is lifted; see e.g. (\ref{eq:Taub-NUT1}) below.

The Coulomb branch of ${\cal N}=4$ SQED
maps to the moduli space spanned by $v_{\pm}$. Classically, one can identify\footnote{This follows from the $U(1)_J$ charges of $V_\pm$, the periodicity of the dual photon $\gamma \to \gamma +g^2$, and the fact that the tree level kinetic term fixes the holomorphic coordinate to be $\sigma + i \gamma$. }
\begin{equation}
\label{mirrormap}
v_{\pm} \sim {\rm exp}\left(\pm 2\pi{\sigma + i\gamma \over g^2}\right)~.
\end{equation}
On the Coulomb branch $\vec \phi\equiv  (\sigma, \frac{1}{\sqrt 2} \,{\rm Re} \phi,\frac{1}{\sqrt 2} \,{\rm Im} \phi)$ the gauge coupling function receives one-loop corrections, 
\be\label{eq:grunning}
\frac{1}{g_L^2}= \frac{1}{g^2}+ \frac{1}{4\pi |\vec \phi|}
\ee
while higher loop and nonperturbative corrections are absent for abelian $\mc N=4$ theories. The quantum-corrected moduli space is given by a sigma model with Taub-NUT metric~\cite{Mirror,deBoer:1996mp}, 
\be\label{eq:Taub-NUT1}
L_\text{eff}= \frac{1}{2g^2} \left(H(\phi)(\partial_\mu \vec \phi)^2 + H^{-1}(\phi) (\partial_\mu \gamma + \frac{1}{2\pi}\vec \omega \cdot \partial_\mu \vec \phi)^2 \right)
\ee
with
\be\label{eq:Taub-NUT2}
H(\phi) = 1 + \frac{g^2}{4\pi | \vec \phi|}\;\;,\;\;\vec \nabla \times \omega = \vec \nabla H\,.
\ee

The IR limit $g^2/|\phi| \to \infty$ of this sigma model is the mirror free hypermultiplet theory. This can be seen by redefining the fields (see e.g.~\cite{Yin})
\be\label{eq:quantum-map}
\left(
\begin{matrix}
v_+ \\ v_-^*
\end{matrix}\right)= \sqrt{\frac{|\vec \phi|}{2\pi}} e^{2\pi i \gamma/g^2}
\left(
\begin{matrix}
\cos \frac{\theta}{2} \\
e^{i \lambda} \sin \frac{\theta}{2}
\end{matrix}
\right)\;\;,\;\;\vec \phi = |\vec \phi| (\cos \theta, \sin \theta \cos \lambda, \sin \theta \sin \lambda)
\ee
which indeed yields the free Lagrangian
\be
L_\text{eff}= |\partial_\mu v_+|^2 + |\partial_\mu v_-|^2\,.
\ee
A similar change of variables gives the fermions in the hypermultiplet. This mirror map between the electric and magnetic theories
will be quite useful later. It will allow us to find the IR fate (in the
dual variables) of solutions of the SQED theory, by mapping them to
exact solutions in the dual.

\subsection{Mirror symmetry for ${\cal N}=2$ theories}\label{subsec:N2mirror}

The ${\cal N}=2$ mirror pair we will consider is a simple modification of the ${\cal N}=4$ example.
As we discussed in \S \ref{subsec:3dsusy}, in the electric theory we can flow from $\mc N=4$ to $\mc N=2$ by adding a chiral superfield $S$ with superpotential interaction $W = S \Phi$. Both $S$ and $\Phi$ become massive, and in the IR one is left with the $\mc N=2$ $U(1)$ gauge theory. This theory has both a Coulomb branch parametrized by $\sigma+2\pi i \gamma$, and a Higgs branch where the meson $M = Q \t Q$ gets an expectation value.

This flow is useful because it allows one to determine the mirror $\mc N=2$ theory. The nontrivial Higgs branch of the electric theory implies that the magnetic theory should contain an additional chiral superfield $M$, besides $V_\pm$.
Since the F-term for $\Phi$ in the electric theory sets $S = Q \t Q$ and the mapping of the moduli space is $\Phi \sim V_+ V_-$, 
the superpotential deformation in the magnetic theory becomes
\begin{equation}
\label{ntwow}
W =h M V_+ V_-\,,
\end{equation}
where $h$ is a coupling with dimensions of mass$^{1/2}$.
Its moduli space of vacua therefore has three branches, depending on which of the (complex scalars
in the) chiral multiplets is non-vanishing.  They meet at an interacting conformal field theory at the
origin, which is a supersymmetric generalization of the Wilson-Fisher fixed point.

\begin{figure}[h!]
\begin{center}
\includegraphics[width=0.4\textwidth]{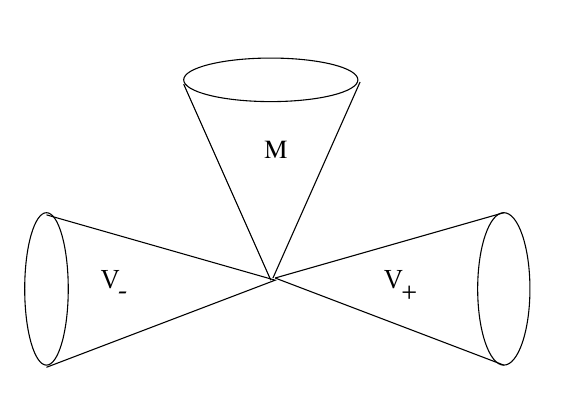}
\end{center}
\caption{\small{The quantum moduli space of the ${\cal N}=2$ SQED theory near the origin agrees with the classical moduli
space of the supersymmetric Wilson-Fisher theory.}}\label{fig:threecones}
\end{figure}

At low energies, or equivalently close to the origin of the moduli space $g^2/|\sigma|\gg 1$, the quantum-corrected SQED coupling grows small according to (\ref{eq:grunning}).  This means that the radius of the dual photon (which is proportional to $g^2$ in the UV) shrinks to zero at the origin of moduli space. Quantum corrections therefore `split' the common meeting locus of the
Higgs branch and Coulomb branch in the gauge theory into a junction between three cones, as in Figure \ref{fig:threecones}. This agrees with the classical moduli space of vacua of the mirror with superpotential (\ref{ntwow}).

Finally, the global symmetries of the electric and magnetic theories are given by
\be
\begin{tabular}{c|ccc}
& $U(1)_R$ & $U(1)_J$ & $U(1)_A$ \\
\hline
$Q$ & 0 & 0 & 1 \\
$\t Q$ & 0 & 0 & 1 \\
$M$ & 0 & 0 & 2 \\
$V_\pm$ & 1 & $\pm 1$ & -1  \\
\end{tabular}
\ee
Here $U(1)_R$ is the $\mc N=2$ R-symmetry, $U(1)_J$ is the topological symmetry $\star F$ of the gauge theory discussed before, and $U(1)_A$ is a global axial symmetry. The dual photon acquires axial charge due to a one-loop BF term, while this effect is seen at tree level in the mirror from (\ref{ntwow}).

Unlike the $\mc N=4$ mirror pair, here we have a strong/strong duality, valid at energy scales much smaller than the relevant couplings of the electric and magnetic theory. While neither side provides a weakly coupled description of the long distance physics, the duality is still physically interesting, connecting a supersymmetric version of the Wilson-Fisher fixed point to a theory with an emergent $U(1)$ gauge field. The mapping between particle and vortex excitations also plays an important role in understanding the dynamics in the presence of defects, to which we turn next.

\section{SUSY defects and mirror symmetry in ${\cal N}=4$ theories}\label{sec:defectsN4}

Now we are ready to consider the addition of external electric or magnetic sources, which amount to turning on finite density and/or magnetic fields. Our first goal is to determine whether these sources can preserve some supersymmetry. We find that it is possible to have finite density or magnetic fields that preserve half of the supercharges. This is an important step, because it allows us to construct and study general (possibly space-dependent) configurations by superposing half BPS pointlike defects. In the second part of our analysis, we will use mirror symmetry to understand the IR dynamics in the presence of supersymmetric defects.
UV sources which interact with a strongly coupled theory backreact on the field configuration in a way which is summarized by the mirror solution. We also discuss how some of our results can be interpreted in terms of insertions of Wilson and 't Hooft line operators. For related work on such operators in the context of mirror symmetry see~\cite{Borokhov:2002cg,Kapustin:2012iw}.

\subsection{Adding electric charges to ${\cal N}=4$ SQED}\label{subsec:electricN4}

We would like to add (static) external charges, with some charge-density $\rho(x)$, to the theory.  We 
accomplish this by adding a  source term for the $U(1)$ gauge field
\begin{equation}
L \supset \frac{1}{2\pi}\rho(x) A_0~.
\end{equation}
However, the charge density by itself breaks supersymmetry: the action is no longer invariant under the SUSY variations (\ref{eq:susyvector}).

Let us view the density as an expectation value for a background gauge field $\hat A_\mu$ that couples to the dynamical $A_\mu$ via a BF interaction $L \supset \epsilon_{\mu\nu\rho} A^\mu \hat F^{\nu\rho}$. Because of the form of this coupling, a charge density is obtained from a background magnetic field,
\be
\rho(x) = \frac{1}{2} \epsilon_{ij}\hat F^{ij}\,,
\ee
a relation that will have important consequences for the mirror description at long distances.
In order to preserve some supersymmetry, we need to turn on additional sources in the background vector superfield $\hat{\mc V}$ (which has $\hat A_\mu$ as one of its components) and allow for the supersymmetrization of the BF interaction, Eq.~(\ref{eq:BF}). Specifically, we add the following source terms to the $\mc N=4$ SQED theory:
\be\label{eq:Lsource1}
L_\text{source} =\frac{1}{2\pi}\left( \frac{1}{2}  A_0\, \epsilon^{ij} \hat F_{ij}(x)+\sigma \hat D(x)\right)\,,
\ee
where $\hat F_{ij}$ and $\hat D$ depend on space but not on time. 

Now, let us imagine weakly gauging $\hat {\mc V}$; supersymmetry will be preserved if the variation of the gaugino $\hat \lambda$ vanishes,
\be
\delta \hat \lambda =\left(\frac{1}{2} \gamma^i \gamma^j \hat F_{ij}(x)- i \hat D(x) \right) \epsilon=0\,.
\ee
Recalling that the gamma matrices in $2+1$ dimensions satisfy (in our conventions)
\be\label{eq:gamma-identity}
\gamma^\mu \gamma^\nu = g^{\mu\nu} \mathbf{1} - i \epsilon^{\mu\nu\rho}\gamma_\rho\,,
\ee
half of the supersymmetries are preserved, $(1 \pm \gamma_0) \epsilon=0$,
as long as
\be\label{eq:susyFD}
\frac{1}{2} \epsilon_{ij} \hat F^{ij}(x) = \pm\, \hat D(x)\,.
\ee
Therefore, it is possible to have supersymmetry at finite density as long as we add the extra source $\hat D$ determined by (\ref{eq:susyFD}).

Another way of proving (\ref{eq:susyFD})  --which does not weakly gauge $\hat {\mc V}$-- is to require that the solutions of the equations of motion in the presence of sources preserve supersymmetry.  Working for simplicity at weak coupling and ignoring the interactions with matter fields,\footnote{This is a good approximation far along the Coulomb branch.} the equations of motion in the presence of the sources are
\begin{equation}
\label{EOM}
{1\over g^2} \partial_i^2 A_0 = \frac{\rho(x)}{2\pi}\;\;,\;\;{1\over g^2}\partial_i^2 \sigma(x) = - \frac{\hat D(x)}{2\pi}~.
\end{equation}
The index $i$ runs over the spatial directions.
Now, supersymmetry requires that the SUSY variations of all fermions should vanish on the
background bosonic field configuration.  The gaugino variation $\delta \lambda=0$ imposes, from (\ref{eq:susyvector}),
\begin{equation}
 (\partial_i A_0 \gamma^0 - \partial_i \sigma) \epsilon =0\,,
\end{equation}
where we have set $D=0$. In order to preserve half of the supercharges, we need to impose, up to a constant,
\begin{equation}
\label{bps}
A_0 = \pm \sigma~,
\end{equation}
with the unbroken supercharges being $(1 \mp \gamma_0) \epsilon = 0$.
From (\ref{bps}) together with (\ref{EOM}), this amounts to a condition on the sources $\hat D= \mp \rho$, which gives indeed (\ref{eq:susyFD}).  This approach is equivalent to the requirement that the Lagrangian is invariant under the SUSY transformations Eq.~(\ref{eq:susyvector}).

In summary, an external charge distribution added to the path integral via an insertion of
\begin{equation}
W_{\rho} = {\rm exp} \left( -\frac{i}{2\pi} \int dt d^2x \rho(x) (A_0 \pm \sigma) \right)
\end{equation}
can preserve half of the supersymmetry.  In the special case where $\rho(x) = \delta^2(x)$, this is
simply the familiar 1/2 BPS Wilson line of supersymmetric gauge theories.

\subsubsection{Classical solutions and field of a point charge}

We can present solutions to the equations of motion in the UV (where $g \to 0$) for rather general sources. The dynamics at long distances is strongly coupled and will be analyzed below using the mirror dual description.

Let us switch for convenience to complex coordinates
\begin{equation}\label{eq:complex}
z = {x_1 + i x_2 \over \sqrt{2}}\,,\,\, \partial_z = {\partial_1 - i\partial_2 \over \sqrt{2}}\,,\,\,A_z = 
{A_1 - iA_2 \over \sqrt{2}}~
\end{equation}
with the obvious definitions for complex conjugates. We recall that in two dimensions
\begin{equation}
\label{GF}
\partial \bar\partial {\rm log}(z \bar z) = 2\pi \delta^2(z,\bar z)~.
\end{equation}

Equipped with this Green's function, we can immediately write down the UV solution generated by a
given charge distribution:
\begin{equation}\label{eq:linear-superposition}
A_0(z,\bar z) = \pm \sigma(z, \bar z) = {g^2 \over 8\pi^2} \int d^2u \,\rho(u,\bar u) {\rm log} \vert
u-z \vert^2~.
\end{equation}
One is free to add a homogeneous solution to (\ref{EOM}), which corresponds to shifting
\begin{equation}
A_0 \to A_0 + f(z) + \bar f(\bar z)
\end{equation}
(where $\bar f$ must be the complex conjugate of $f$ to keep the gauge field real).  Boundary
conditions at infinity can determine the homogeneous solution.  For instance, for a 
spherically symmetric distribution of external charges, we would wish to find a spherically
symmetric solution and set $f(z) = 0$.

The dual photon also has interesting behavior.  From its definition (\ref{eq:dualphoton}), it follows that
$\partial_z \gamma = i\partial_z A_0,~\partial_{\bar z}\gamma = -i \partial_{\bar z} A_0$.
The solution to these equations is
\begin{equation}
\gamma(z,\bar z) = i {g^2\over 8\pi^2} \int d^2u ~\rho(u,\bar u) {\rm log}({u-z \over {\bar u - \bar z}})
~.
\end{equation}
On the Coulomb branch, the $\sigma$ scalar in the gauge supermultiplet combines with $\gamma$ to form a complex scalar, whose solution is then
\begin{equation}
\sigma + i \gamma = {g^2\over 8\pi^2} \int d^2 u ~ \rho(u,\bar u)\left[ \pm {\rm log}|u-z|^2 -
{\rm log}{u-z \over {\bar u - \bar z}}\right]~.
\end{equation}
Therefore supersymmetric solutions have either a holomorphic or
antiholomorphic profile for $\sigma + i \gamma$.

As the simplest example, consider a point-like external charge,
\begin{equation}
\rho(z,\bar z) = 2\pi q_0 \,\delta^2(z,\bar z)~.
\end{equation}
The gauge field and $\sigma$ take the form
\begin{equation}
A_0(z, \bar z) = \pm \sigma(z, \bar z) = -\frac{1}{4\pi}g^2 q_0 ~{\rm log} \frac{ |z|^2}{r_0^2}\,,
\end{equation}
where the constant $r_0$, added for dimensional reasons, is related to the Coulomb branch expectation value.
The dual photon becomes
\begin{equation}
\gamma = -g^2 q_0 \,\frac{\theta}{2\pi}
\end{equation}
and $\theta$ is the standard angular variable on the complex plane, $z = r e^{i\theta}$. Note that when going around an electron of unit electric charge the dual photon has a monodromy $\gamma \to \gamma + g^2$ , which equals its periodicity.

Of course, the profile of the electric potential for external charges in the non-interacting limit is well known. However, we will next consider the long distance behavior, and find a much more interesting solution. Surprisingly,
strong screening effects from the chiral multiplets $(Q, \t Q)$ turn the logarithmic running into a constant. Also, 
the linear superposition (\ref{eq:linear-superposition}) will flow to a product of monomials, each of which can be interpreted as a localized vortex for the topological $U(1)_J$.



\subsection{Global vortices in the mirror configuration}\label{subsec:global-vorticesN4}

As one flows to the IR in the ${\cal N}=4$ SQED theory, $g \to \infty$ and the magnetic description in terms of the free hypermultiplet $V_\pm$ is more appropriate.  We will use the mirror map, reviewed in \S \ref{subsec:mirrorN4review}, to find the long distance dynamics of the supersymmetric finite density configuration. Let us first directly study this magnetic theory and then map it to the electric variables.

According to (\ref{eq:mirror-map1}), an external charge density in the electric theory corresponds to a $U(1)_J$ background magnetic field in the mirror dual, while the electric source term $\hat D$ is a background D-term. As before, the sources can have arbitrary dependence on space, but need to be static.
The Lagrangian for the bosonic fields
takes the form
\begin{equation}
\label{magneticlag}
L = |\hat D_{\mu} v_+|^2  + |\hat D_\mu v_-|^2 + \hat D(x)(|v_+|^2 - |v_-|^2)~,
\end{equation}
where $\hat D_\mu v_{\pm} = (\partial_{\mu} \pm i \hat A_{\mu}) v_{\pm}$.

We wish to find supersymmetric configurations with a non-zero static magnetic field $\hat B = \hat F_{12}$,
or in complex coordinates
\begin{equation}\label{eq:gauge-potential1}
\partial_z \hat A_{\bar z} - \partial_{\bar z} \hat A_z = i\hat B(z,\bar z)~.
\end{equation}
Weakly gauging the global symmetry and setting to zero the SUSY variation of the gaugino $\delta \hat \lambda=0$ gives, as in (\ref{eq:susyFD}),\footnote{This approach was used in e.g. \cite{Polchinski} in the context of supersymmetric Landau levels.} 
\be\label{eq:susyBD}
\hat B(x) = \pm \hat D(x)\;,\;(1\pm \gamma_0)\epsilon=0\,.
\ee
However, since the magnetic theory is free, it is more instructive to follow a different route, which will also yield the first order equations that need to be satisfied by $v_\pm$.

Unbroken supersymmetry requires\footnote{According to our conventions (\ref{eq:complex}), $\gamma^z = \frac{1}{\sqrt{2}}(\gamma^1 + i\gamma^2)$, $D_z = \partial_z + i e A_z$,  $D_{\bar z} = \partial_{\bar z} + i e A_{\bar z}$, etc.}
\begin{equation}
\delta\psi_{\pm} = -i (\gamma^z D_z v_{\pm} + \gamma^{\bar z}D_{\bar z}v_{\pm})\epsilon = 0~.
\end{equation}
Half of the supersymmetries are preserved if
\begin{equation}
\label{topcase}
\gamma^z \epsilon = 0, ~~D_{\bar z} v_{\pm} = 0
\end{equation}
or
\begin{equation}
\label{bottomcase}
\gamma^{\bar z} \epsilon = 0,~~D_{z} v_{\pm} = 0~.
\end{equation}

On the other hand,
the equations of motion following from (\ref{magneticlag}) are
\begin{equation}
\label{lleqn}
(D_z D_{\bar z} + D_{\bar z}D_z) v_{\pm} \pm \hat D(z,\bar z) v_{\pm} = 0\,.
\end{equation}
Let us choose the SUSY case (\ref{topcase}), for which $D_{\bar z} v_{\pm} = 0$. Plugging into (\ref{lleqn}), and since in the presence of the external $\hat B$ field,
\begin{equation}
[ D_z, D_{\bar z}] v_{\pm} = \mp \hat B(z,\bar z) v_{\pm}\,,
\end{equation}
we find that the equation of motion can be solved if
\begin{equation}
\hat B(z,\bar z) = -\hat D(z,\bar z)~.
\end{equation}
In the other case (\ref{bottomcase}) one instead requires $\hat B = \hat D$.  This reproduces (\ref{eq:susyBD}).

We conclude that, in the presence of an external magnetic field (with arbitrary space dependence), the free hypermultiplet theory admits solutions that preserve half of the supercharges if an external D-term (\ref{eq:susyBD}) is also turned on. 

\subsubsection{Classical solutions and external vortices}

Let us now study the supersymmetric classical solutions in the presence of an external magnetic field and D-term $\hat D(x) = \pm \hat B(x)$.

Unlike the electric theory where only the field strength $\hat F_{\mu\nu}$ appears, here the hypermultiplet couples to the background potential $\hat A_\mu$, so we need to choose a gauge. Given an external magnetic field $\hat B$, a convenient choice that solves (\ref{eq:gauge-potential1}) is
\begin{equation}
\label{genA}
\hat A_z = -{i \over 4\pi} \partial_z \int d^2u ~\hat B(u,\bar u) {\rm log} |u-z|^2,~
\hat A_{\bar z} = {i \over 4\pi} \partial_{\bar z} \int d^2u ~\hat B(u,\bar u) {\rm log} |u-z|^2\,,
\end{equation}
which uses the Green's function on the plane.

With $\hat B =- \hat D$, the supersymmetric configurations satisfy $D_{\bar z} v_{\pm} = 0$, (\ref{topcase}).  Plugging in the background gauge field (\ref{genA}), it is straightforward to integrate this equation to obtain
\begin{equation}\label{eq:Dbarzsol}
v_{\pm}(z,\bar z) = f_{\pm}(z) {\rm exp}\left( \pm {1\over 4\pi}\int d^2u ~\hat B(u,\bar u) 
{\rm log}|u-z|^2\right)~,
\end{equation}
with $f_{\pm}$ arbitrary holomorphic functions.  Similarly, with $\hat B =  \hat D$, the solutions to $D_z v_{\pm} = 0$ are
\begin{equation}\label{eq:Dzsol}
v_{\pm}(z,\bar z) = \bar f_{\pm}(\bar z) {\rm exp}\left( \mp{1\over 4\pi} \int d^2u
~\hat B(u,\bar u) {\rm log} |u-z|^2 \right)~.
\end{equation}
These arbitrary (anti)holomorphic prefactors correspond to the gauge freedom noted above, while a constant prefactor moves the configuration along the Higgs branch of the hypermultiplet theory. In order to determine the physically correct choice of $f_\pm$, it is instructive to first discuss pointlike magnetic flux insertions, from which we can then construct a general $\hat B$ by superposition.

Choosing a delta-function localized magnetic flux
\begin{equation}\label{eq:pointlikeB}
\hat B(z, \bar z) = 2\pi q_0 \delta^2 (z, \bar z)~,
\end{equation}
(\ref{genA}) gives simply $\hat A= q_0 d\theta$.
Any loop encircling the insertion at $r=0$ will have a fixed holonomy
$e^{i\oint \hat A} = e^{2\pi i q_0}$, which is the definition of a global vortex.
Every charged field picks an Aharonov-Bohm phase given by the magnetic flux times its charge. This fixes the previous arbitrary prefactor to $f_\pm \propto z^{\mp q_0}$, obtaining the solution
\be\label{eq:localizedvpm}
v_\pm = v_0^{\pm}\left(\frac{\bar z}{ z} \right)^{\pm q_0/2}=v_0^\pm\, e^{\mp i q_0 \theta}
\ee
for $D_{\bar z} v_\pm=0$, and the complex conjugate of (\ref{eq:localizedvpm}) for the SUSY case $D_z v_\pm=0$. The constant $v_0^\pm$ parametrizes the Higgs branch position.

Eqs.~(\ref{eq:pointlikeB}) and (\ref{eq:localizedvpm}) represent an external (nondynamical) pointlike vortex. Let's compare this to an Abrikosov vortex. Recall that the abelian Higgs model in $2+1$ dimensions admits dynamical vortices where the behavior of the scalar field at infinity is $\phi \to v e^{i N \theta}$, with $v$ the minimum of the potential and $N$ the winding number. At long distances, $ A \sim N d \theta$ and the magnetic flux is proportional to $N$. The vortex has a core where the scalar and gauge field vanish. In the present case, the magnetic flux is external and delta-function localized. The scalars in the hypermultiplet have a Higgs branch parametrized by $v_0^\pm $ above, and the effect of the nondynamical vortex is to introduce a nonzero winding, (\ref{eq:localizedvpm}), which is singular at the location of the pointlike core.

Returning now to the general case, the previous discussion fixes $f_\pm$ to give a monodromy determined by the total magnetic flux,
\be
f_\pm(z) = v_0^\pm \exp \left(\mp \frac{1}{2\pi} \int d^2u\,\hat B(u, \bar u)\,\log(u-z) \right)\,.
\ee
In conclusion, the general supersymmetric configurations that reproduce the correct monodromies around the vortex insertions are given by
\be
\label{richsoln1}
\hat B = \hat D\,,\;(1+\gamma_0)\epsilon=0\;,\;D_z v_\pm=0\;,\;v_{\pm} = v_0\, {\rm exp} \left( \pm {1\over 4\pi} \int d^2u
~\hat B(u,\bar u) {\rm log}\frac{u-z}{\bar u - \bar z}\right)~,\nonumber \\
\ee
and
\be
\label{richsoln2}
\hat B = - \hat D\,,\;(1-\gamma_0)\epsilon=0\;,\;D_{\bar z} v_\pm=0\;,\,v_{\pm} = v_0\, {\rm exp} \left( \mp {1\over 4\pi} \int d^2u ~\hat B(u,\bar u)
{\rm log}\frac{u-z}{\bar u - \bar z}\right)~.
\ee

\subsubsection{Mapping to the electric variables}

Finally, consider the map from the electric to the magnetic theory given by (\ref{eq:quantum-map}), a result which is exact in the regime $|\vec \phi|/g^2 \ll 1$. The mapping of external sources is $\rho(x) = \hat B$, and $\hat D$ is the same on both sides. This means that an external electron of the SQED theory maps to a global $U(1)_J$ vortex in the magnetic theory. For a pointlike charge $\rho = 2\pi q_0 \delta^2(z, \bar z)$, the electric theory dual photon has monodromy $g^2 q_0$. In the magnetic theory, this effect appears as the Aharonov-Bohm phase of the charged $v_\pm$. Therefore, the phases of the map (\ref{eq:quantum-map}) agree, as they should.\footnote{In both theories, there is a conserved $U(1)$ symmetry from a combination of spatial rotations and a $U(1)_J$ transformation. With the $f_\pm$ found before, the same linear combination is preserved on both sides of the duality.} The duality between
a Wilson line insertion in the SQED theory and a vortex in the magnetic dual is part of the particle/vortex duality in mirror symmetry and has been made more precise in e.g.~\cite{Borokhov:2002cg,Kapustin:2012iw}.

On the other hand, $\sigma$ has a much more interesting behavior at different scales. For a point charge, $\sigma$ diverges logarithmically in the electric theory. However, the behavior at long distances is given by (\ref{eq:quantum-map}),
\be
|\sigma |= 2\pi |v_\pm|^2\,,
\ee
where for simplicity we have set the remaining Coulomb branch coordinates to zero. Recalling (\ref{eq:localizedvpm}), we see that $|\sigma|$ flows to a constant in the IR. The logarithmic behavior has been screened by the strong dynamics of $(Q, \t Q)$ in the electric theory, in a way that is captured by the classical solution of the magnetic dual theory.

The mapping with multiple local sources (or a more general spatially dependent distribution) is also very interesting.
In the electric theory, for a
solution with $\rho(x) = \rho_1(x) + \rho_2(x)$, in the UV $g \to 0$ limit, the solutions are 
additive -- one simply uses the Green's function to add up the contributions due to the two
localized sources.  

By the time one flows to the IR, we see that the superposition of UV sources has created considerably
more complicated effects.  $\rho(x)$ maps directly to $\hat B(x)$ under mirror symmetry.
The solution created by a superposition $\hat B(x) = \hat B_1(x) + \hat B_2(x)$ (with each 
$\hat B_i$ resulting from the localized source in the electric theory) is the exponential 
(\ref{richsoln1}), exhibiting additivity in the exponent and a very complicated interaction between
the sources!  The field configurations (\ref{richsoln1}) and (\ref{richsoln2}) capture the complicated process (due to strong dynamics) by which the field configuration sourced by external charges in ${\cal N}=4$ SQED changes as one flows to the IR.
We will use this in \S \ref{sec:lessons} to solve for the behavior of defects in ${\cal N}=4$ SQED as one flows to strong coupling.

\subsection{Adding magnetic charges to ${\cal N}=4$ SQED}\label{subsec:magneticN4}

Now we consider the reversed situation, where we turn on a source for the $U(1)$ magnetic field of the SQED theory. In the mirror dual, this corresponds to a chemical potential for $U(1)_J$. We will first study the conditions under which this can be done in a supersymmetric fashion before analyzing the solutions in the presence of spatially dependent sources. An important difference with the previous situation is that now the theory will admit BPS configurations of finite central charge, and this will allow us to understand various aspects of the correspondence between particles and vortices across the duality.

As before, the sources are part of the $\mc N=4$ background vector multiplet $\hat{\mc V}$, which appears in the terms (\ref{eq:BF}). A source for $\partial_i A_j$ is given by a nonzero $\hat A_0(x)$.  Following the steps of the previous sections, we weakly gauge $\hat{\mc V}$ and impose $\delta \hat \lambda=0$, Eq.~(\ref{eq:susyvector}). This shows that the other source that needs to be turned on in order to preserve SUSY is the background scalar $\hat \sigma$, which plays the role of an FI term for the $U(1)$ gauge theory. We study the SQED theory in the presence of
\be\label{eq:sourcehatA0}
L_\text{source}= \frac{1}{2\pi} \left(\hat A_0(x)\,\epsilon_{ij} \partial_i A_j+ \hat \sigma(x) D \right)\,.
\ee
The solution to $(\gamma^0 \partial_i \hat A_0 - \partial_i \hat \sigma) \epsilon=0$ is
\be\label{eq:susyA0}
\hat A_0 (x) = \pm \hat \sigma(x)\;\;,\;\;(1 \mp \gamma_0)\epsilon=0\,.
\ee

At this stage, a somewhat subtle point needs to be addressed. Eq.~(\ref{eq:BF}) defined the BF interaction to be of the form $\epsilon^{\mu\nu\rho} \hat A_\mu  F_{\nu\rho}$, namely a coupling of the topological current $\star F$ to an external gauge field. Furthermore, this form of interaction is explicitly gauge invariant. We could have chosen instead an interaction $\epsilon^{\mu\nu\rho}  A_\mu  \hat F_{\nu\rho}$, differing from the previous one by a boundary term. The equations of motion are not modified, and the boundary term vanishes if the external source falls off fast enough at infinity. However, we will be interested in the possibility of constant FI terms; then $\hat A_0$ and the boundary term do not vanish at infinity. The correct form in such cases is (\ref{eq:sourcehatA0}). This has important consequences for the vacuum structure of the theory, as we discuss shortly.

Let us now turn to the field configurations; we will see that (\ref{eq:susyA0}) also follows from requiring that the solutions to the equations of motion preserve SUSY. We will consider first the theory at the origin of the Coulomb branch. In the presence of the background (\ref{eq:sourcehatA0}), the dynamical magnetic field $B=F_{12}$ and D-term will be turned on. The gaugino variation $\delta \lambda=0$ is satisfied for
\be\label{BDsusy}
\frac{1}{2} \epsilon_{ij} F^{ij}= \pm D\;\;,\;\;(1\pm\gamma_0)\epsilon=0\,,
\ee
where the auxiliary field is
\be
D=-g^2 \left(|q|^2 - |\t q|^2+ \frac{\hat \sigma(x)}{2\pi} \right)\,.
\ee
Furthermore, the conditions $\delta \psi_q = \delta \psi_{\t q}=0$ that preserve half of the supercharges are
\be\label{eq:qsusy1}
D_{\bar z} q = D_{\bar z} \t q =0\;\;,\;\;\gamma^z \epsilon=0
\ee
or
\be\label{eq:qsusy2}
D_z q = D_z \t q =0\;\;,\;\;\gamma^{\bar z} \epsilon=0\,.
\ee

Next, compare these conditions with the classical equations of motion. The $(q, \t q)$ equations are solved for (\ref{eq:qsusy1}) if $B=-D$, while the other case (\ref{eq:qsusy2}) needs $B=D$. These are consistent with the gauge field equation only if $\hat A_0 = \pm \hat \sigma$, agreeing with what we found in (\ref{eq:susyA0}). To summarize, the SUSY conditions at the origin of the Coulomb branch are
\bea\label{eq:vortex-cond}
&&\hat A_0(x) =- \hat \sigma(x)\;,\;B= D\;,\;D_z q= D_z \t q=0 \nonumber\\
&&\hat A_0(x) = \hat \sigma(x)\;,\;B= -D\;,\;D_{\bar z} q= D_{\bar z} \t q=0\,.
\eea
The F-term for $\Phi$ requires $q \t q=0$, so one of the scalars has to vanish. 
The solutions (if they exist) preserve half of the supercharges. These would be generalizations of the BPS vortices to the case of an $x$-dependent FI term and source $\hat A_0$. It would be interesting to determine whether such solutions exist, and study their dynamics.

In the case where the FI term $\hat \sigma$ is a constant, the nontrivial BPS solutions to these
equations are well known.  They are the familiar vortices of the abelian Higgs model, studied in detail in
supersymmetric theories in e.g. \cite{Edelstein,Tong}.  
In a background with $\int d^2x~B = k$
(i.e. $k$ units of magnetic flux), these vortex solutions have a moduli space ${\cal M}_k$ which is $k$ (complex)
dimensional.  One can think of it as being spanned by the positions of the $k$ vortices in the plane, and having
the asymptotic structure ${\mathbb C}^k/S_k$ when the vortices are well separated.

There is, however, an important difference with the SUSY Higgs model: the sources (\ref{eq:sourcehatA0}) include, besides the FI term, a coupling $\hat A_0 B$ to the magnetic field. For constant $\hat A_0$ this term is a total derivative and so the equations of motion are not modified. However, taking into account the SUSY conditions, it gives a negative contribution to the total energy. Due to this effect, the Coulomb branch is not lifted. Indeed,
one can show that
\be\label{eq:vacuum-vortex}
q = \t q =0\;,\;B=\pm g^2 \,\frac{\hat \sigma}{2\pi}\,,
\ee
solve the SUSY variations and equations of motion. This vacuum allows for Coulomb branch expectation values, along which the vortices are lifted. This should be constrasted with the theory with just
an FI term and no $\hat A_0$, for which there is no Coulomb branch.\footnote{Finite energy requires a Higgs branch vacuum $|q|^2 - |\t q|^2 + \hat \sigma/(2\pi) \to 0$ at infinity.} Therefore, our supersymmetric extension of the FI term admits both vortex solutions and  a nontrivial Coulomb branch, without having to turn off the FI term. This will provide a more precise mapping between vortices and particles in the mirror dual, to which we turn next.

\subsection{Finite chemical potential in the mirror dual}\label{subsec:chemical-potential-magnetic}

The magnetic dual involves turning on a chemical potential for $U(1)_J$, as well as a real mass
$\hat \sigma$ for the chiral multiplets $V_{\pm}$. The Lagrangian is
\begin{equation}
\label{magneticlagtwo}
L = |\hat D_\mu v_{\pm}|^2+ i \bar \psi_\pm  \not   \!\!\hat D \psi_\pm - |\hat \sigma|^2 (|v_+|^2 + |v_-|^2)- \hat \sigma (\bar \psi_+ \psi_+- \bar \psi_- \psi_-)
\end{equation}

The supersymmetry variations of the fermions are now (for constant $v_\pm$)
\be
\delta \psi_{\pm} = -\left( i \not   \!\!\hat D v_{\pm} \pm  \hat \sigma v_{\pm}\right)\epsilon=\pm\, v_\pm (\gamma^0 \hat A_0 - \hat \sigma)\epsilon
\ee
Unbroken supersymmetry requires, unsurprisingly, that $\hat A_0 = \pm \hat \sigma$,
with half of the supersymmetry being preserved when this is the case. With this relation between the sources, the positive scalar mass squared from $\hat \sigma^2$ exactly cancels the negative contribution from the chemical potential, leaving
\be\label{eq:susy-electrons}
L = |\partial_\mu v_\pm|^2 + i \bar \psi_\pm \not \! \partial \psi_\pm \mp i \hat A_0 (v_\pm^* \partial_0 v_\pm - v_\pm \partial_0 v_\pm^*)\mp \hat A_0\,\bar \psi_\pm (1+\alpha \gamma^0)\psi_\pm\,,
\ee
with $\alpha = \pm$ the relative sign between $\hat A_0$ and $\hat \sigma$.

Unlike the case with only a chemical potential (which leads to tachyonic scalars), here there is a Higgs branch moduli space where $v_+$ and $v_-$ have arbitrary constant values. Therefore the vacuum is a superfluid where the order parameter for symmetry breaking is not fixed. This is the mirror dual of the Coulomb branch that appears for (\ref{eq:vacuum-vortex}). 

Let us discuss in more detail the particle excitations in the presence of a constant background $\hat \sigma$ field and the chemical potential. Since $(1\pm \gamma_0)/2$ is a projector on the spinor indices, the background gives
half of the fermion components in each $\psi_\pm$ a mass  $2 \hat A_0$, while the other half are massless.
Intuitively, the real mass $\hat \sigma$ in the absence of a chemical potential would give a mass to 
all of the particles.  The presence of the chemical potential decreases the mass of the antiparticles to zero, and
increases the mass of the particles (or the reversed, depending on the sign of $\hat A_0$).
Similarly, the scalar equation of motion in this class of backgrounds becomes
\begin{equation}
\Box v_\pm= \pm 2 i\hat A_0 \,\partial_0 v_\pm\,,
\end{equation}
which admits plane-wave solutions of frequency $\omega = \hat A_0 \pm \sqrt{\hat A_0^2 + {\bf k}^2}$
for $v_+$, and a similar solution for $v_-$ with $\hat A_0 \to - \hat A_0$. The bosons and fermions have degenerate masses, as expected from supersymmetry. The massless root describes an infinitesimal fluctuation along the Higgs branch, while
the massive root, with $\omega = 2\hat A_0$ at ${\bf k}=0$, corresponds to a massive BPS particle.

We are now in a position to describe the mirror configuration to $k$ BPS vortices.  Because the $U(1)_J$ 
current is $\star F$, a magnetic field for the dynamical gauge field maps to the presence of $U(1)_J$ charges in the free hypermultiplet
description. The mirror configuration is then precisely $k$ of the massive BPS particles discussed above.
The moduli space of the $k$ positions of these particles maps to the moduli space of $k$ abelian vortices
in the SQED description (though the detailed geometry of the moduli space is expected to be different).  Furthermore, the superfluid phase where the $U(1)_J$ is broken along the Higgs branch $(v_+, v_-)$ maps to the Coulomb branch, where the vortices in the electric theory are lifted. We can interpolate smoothly between the isolated vacua with vortices and the Coulomb branch without changing the BPS spectrum or external sources, a phenomenon which is manifest classically in the free hypermultiplet dual.
This gives a very explicit realization of the particle/vortex duality.

\section{SUSY defects and mirror symmetry in ${\cal N}=2$ theories}\label{sec:defectsN2}

We now discuss the dynamics of defects in the $\mc N=2$ theories of \S \ref{subsec:N2mirror}. In three dimensions, this is the smallest amount of supersymmetry for which there exist SUSY defects; the reason is that our mechanism needs a vector multiplet (either gauge or global) that contains both a scalar and a gauge field.

\subsection{Semiclassical solutions}

At the semiclassical level, the story with 3d ${\cal N}=2$ supersymmetry is quite similar to the $\mc N=4$ theory. The main difference is that the electric theory has no analog of the chiral superfield $\Phi$ (and hence $W = \sqrt{2} Q \Phi \t Q$ is absent), while the magnetic theory has a new singlet superfield $M$ with a superpotential coupling $W = h M V_+ V_-$. As a consequence, 
there are new branches in the moduli space of vacua
-- the Higgs branch with $q, \tilde q \neq 0$ in the ${\cal N}=2$ SQED theory, and
the (mirror) M-branch in the ${\cal N}=2$ Wilson-Fisher theory.

The semiclassical story for the SUSY defects then carries over as follows. Consider first adding electric charges to the SQED theory. The classical solutions in the UV are the same as in \S \ref{subsec:electricN4}; the field $\Phi$ didn't play an important role for the $\mc N=4$ defects (which introduced sources for $A_0$ and $\sigma$), and is absent in the $\mc N=2$ case. The magnetic dual has nonzero $U(1)_J$ magnetic field and D-term backgrounds, and the solutions are global vortices of the type described in \S \ref{subsec:global-vorticesN4}. However, unlike the $\mc N=4$ case, now the F-term for $M$ forces either $v_+$ or $v_-$ to vanish. This maps to the statement that the field $\Phi$ is absent from the electric theory. Furthermore, $M=0$ along the $v_+$ or $v_-$ branches. 

The mapping of the $M$--branch is more nontrivial. Note that in the magnetic theory we can set $v_+=v_-=0$ because the backgrounds only multiply quadratic functions of $v_\pm$. This should be contrasted with the electric theory, where $\sigma$ and the dual photon acquire spatial dependence and cannot vanish for generic sources. At the origin of the $v$--branches, the $M$--branch opens up. What happens in the electric theory is that $q$ and $\t q$ obtain positive masses proportional to $\sigma^2$, but these are exactly cancelled by the negative contribution from $A_0^2$.\footnote{This is similar to what we found in (\ref{eq:susy-electrons}), but for a dynamical gauge field instead of a background.} Therefore, even though $\sigma$ is nonzero, $(q, \t q)$ can have arbitrary expectation values subject to the D--flatness condition $|q|=|\t q|$, in agreement with the dimension of the $M$--branch.

Next, let us briefly discuss SQED theory in the presence of magnetic sources, the $\mc N=2$ analog of \S \ref{subsec:magneticN4}. Now  $q \t q=0$ is no longer an F-term condition, so $q$ and $\t q$ can be varied keeping $|q|^2 - |\t q|^2$ fixed. This maps to the modulus $M$ of the mirror dual. Also, as explained before, the absence of the field $\phi$ in the electric theory maps to the F-term $v_+ v_-=0$ in the dual. Taking these differences into account, the classical solutions are the same as in the $\mc N=4$ case.

\subsection{Comments on quantum dynamics}

The quantum version of the $\mc N=2$ case is much richer than the $\mc N=4$ one, with both electric and magnetic theories flowing to strong coupling in the IR. Both become equivalent for energies $E \ll g^2$ and $E \ll |h|^2$, where $g$ is the SQED gauge coupling and $h$ is the superpotential coupling (\ref{ntwow}). The K\"ahler potential in $\mc N=2$ theories is not protected, so physical couplings and correlation functions receive large quantum corrections on both sides.

The external electric or magnetic sources require $A_\mu$ and $\sigma$ in the electric theory to be nonzero, and similarly $v_\pm$ are turned on in the magnetic dual. These expectation values need to be much smaller than the respective relevant couplings in order for the theories to be dual. This is the regime of strong coupling, and the semiclassical solutions discussed before will receive important quantum corrections. With the amount of supersymmetry preserved by the defects (two supercharges), the expectation values of fields in the presence of sources cannot be obtained analytically. In particular, the physical gauge coupling, which determines the Coulomb
branch metric, receives higher loop corrections in $\mc N=2$ theories, in contrast with (\ref{eq:grunning}).

In some respects, this situation is similar to the nonsupersymmetric duality between the $U(1)$ abelian Higgs model and the XY model
of~\cite{Peskin,Dasgupta}, where both sides are strongly coupled. Some of the possible IR phases at nonzero chemical potential were discussed in~\cite{SachdevCompr}. There are, however, important differences between the two systems. In the $\mc N=2$ duality discussed here there are flat directions that are protected by supersymmetry, while these are absent in the nonsupersymmetric version. For example, in the magnetic theory doped with $U(1)_J$ chemical potential in a supersymmetric fashion, we obtained a superfluid phase with arbitrary expectation values $\langle v_\pm \rangle$. This feature is expected to survive even at strong coupling, at least if the external sources are localized within a finite region in space. Another important distinction with the nonsupersymmetric case is the existence of BPS excitations in the presence of external electric and magnetic sources. Because of the additional control from supersymmetry it may be possible to analyze to some extent the IR dynamics of doped $\mc N=2$ theories, a question which we hope to return to in the future.

\section{Lessons for impurity physics}\label{sec:lessons}

Mirror symmetry can be a powerful tool for elucidating the physics of impurities, which is of significant interest in condensed matter physics (see e.g. \cite{Buragohain,impure}). This is clearest in the duality between
${\cal N}=4$ QED and the free ${\cal N}=4$ hypermultiplet. For instance, we may start with an arbitrary array of impurities in the electric theory (represented by a configuration of localized sources in $\rho(x)$); map the
sources to the magnetic theory via the mirror map $\hat B = \rho$; and then find the magnetic solution corresponding
to the sources using (\ref{richsoln1}).

The physics of the IR solution in the electric theory can then be found by applying the map between variables
\begin{equation}
\label{mapagain}
v_\pm = \sqrt{\vert \sigma \vert \over 2\pi} e^{\pm 2\pi i \gamma / g^2} ~
\end{equation}
(appropriate for solutions with $\theta = 0$ in \S \ref{subsec:mirrorN4review}).
It is notable that the resulting solution of the electric theory, applicable in the IR, is dramatically different from the naive solution one would obtain using the Green's functions of the classical UV theory to superpose effects of the defect sources. We illustrate this with several examples now.

\begin{figure}[ht!]
\begin{center}
\includegraphics[width=0.6\textwidth]{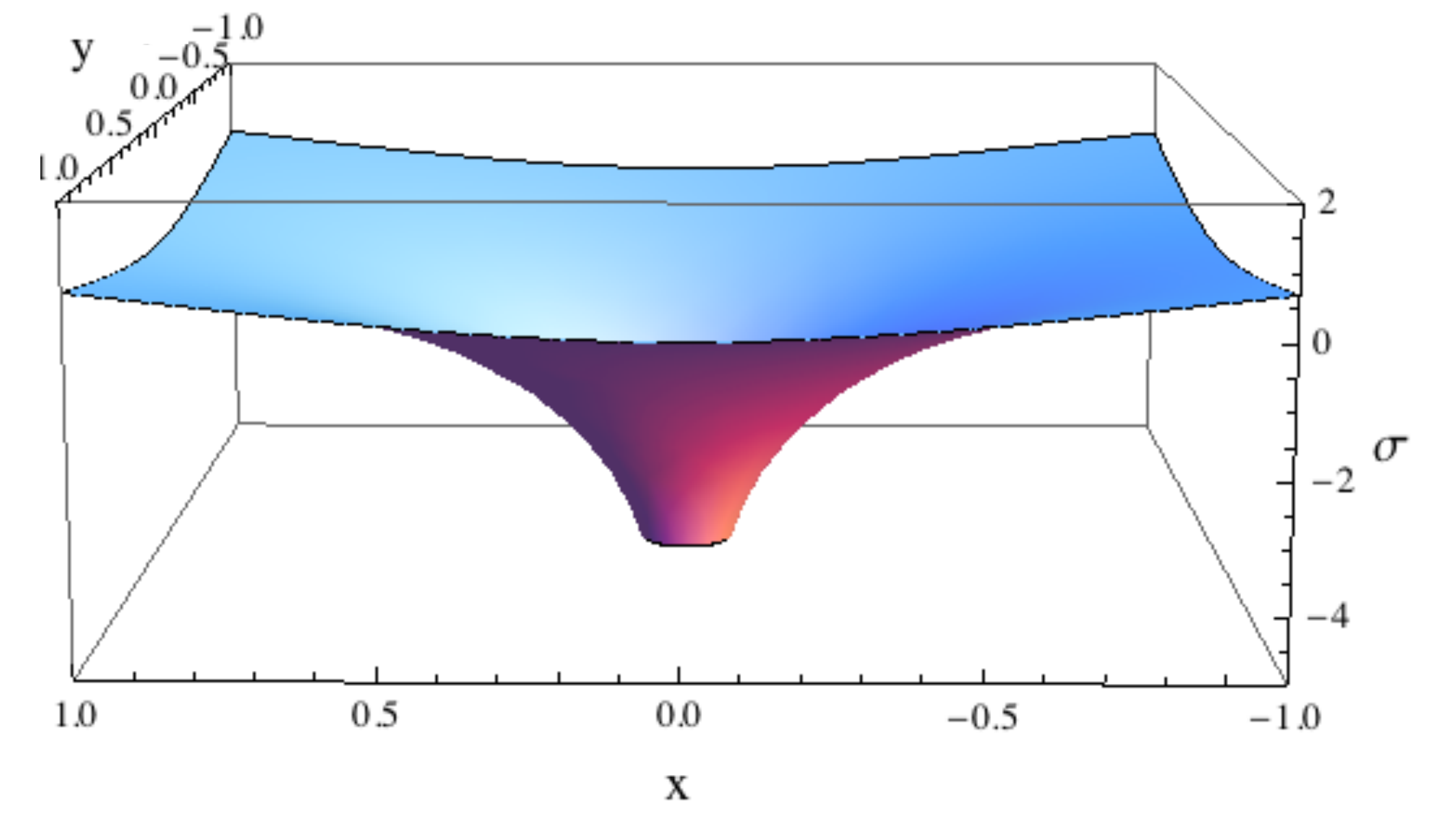}
\end{center}
\caption{\small{The solution for a single electrically charged defect in the UV limit of SQED.}}
\label{fig:electric-single-defect}
\end{figure}

\begin{figure}[ht!]
\begin{center}
\includegraphics[width=0.6\textwidth]{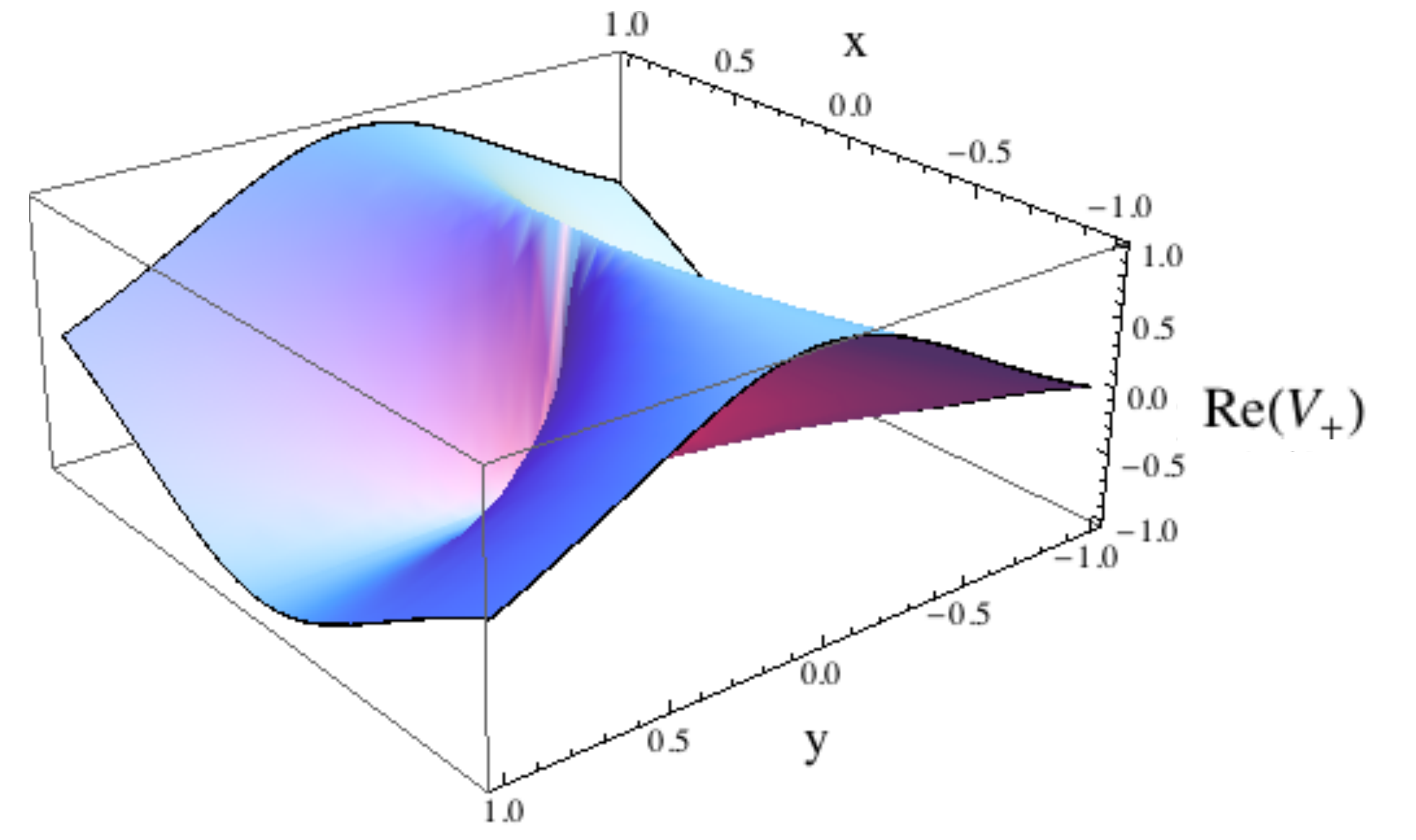}
\end{center}
\caption{\small{The backreaction of a single `external vortex' in the magnetic theory.}}
\label{fig:magnetic-single-defect}
\end{figure}

We start with a depiction of the basic electric defect in Figure \ref{fig:electric-single-defect}. This is the solution from \S \ref{subsec:electricN4} with
\begin{equation}
A_0 =\pm \sigma =- \frac{g^2}{4\pi} \,q_0\, {\rm log}|z|^2~.
\end{equation}
This isolated defect is mapped in the magnetic theory to $v_\pm= v_0^\pm e^{\pm i q_0 \theta}$.
A plot of the single vortex appears in Figure \ref{fig:magnetic-single-defect}.
The constant coefficient $v_0^\pm$ determines which point on the moduli space of vacua one approaches asymptotically. More precisely, for an external vortex/anti-vortex pair, the winding at infinity vanishes, and one approaches some well defined point in moduli space both in the electric and magnetic theories.

Mapping the single external vortex back to the IR electric theory using (\ref{mapagain}), one finds an extremely simple dressed solution,
$\sigma \sim {\rm const},~\gamma \sim {\rm arg}(z)$.
The dramatic change from ${\rm log}$ growth to constant behavior visible in the dressed solution is an effect
of screening of the defect charge by the vacuum polarization of the strongly coupled SQED theory.

\begin{figure}[ht!]
\begin{center}
\includegraphics[width=0.7\textwidth]{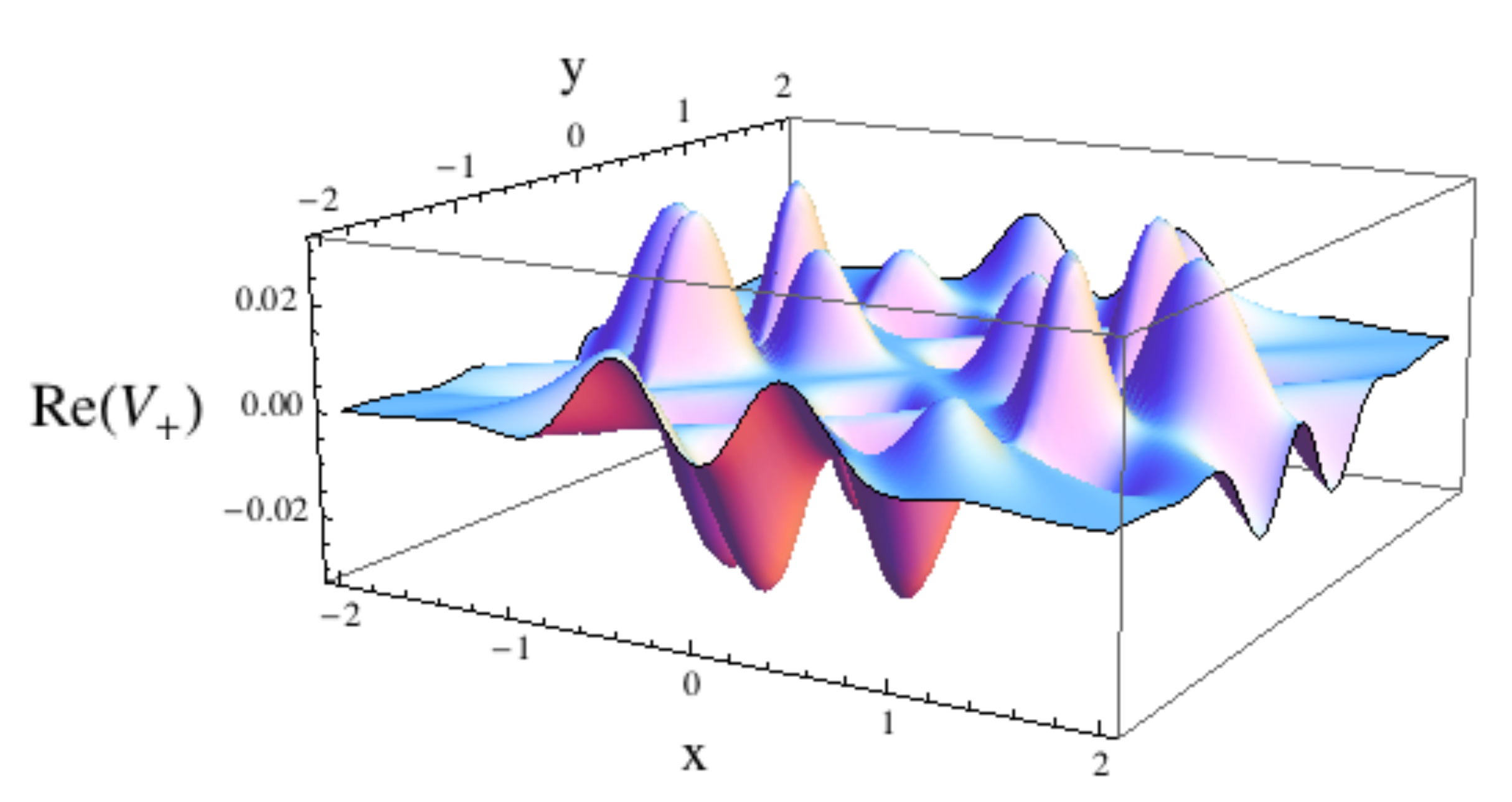}
\end{center}
\caption{\small{A 3$\times$3 lattice of external vortices.}}
\label{fig:threebythree}
\end{figure}

We can use this same technique to find the solution for an arbitrary array of $\delta$-function localized defects, such as
a defect lattice. With localized sources at positions $z_i$ in the spatial plane, the UV electric theory
solution is
\begin{equation}
A_0(z,\bar z) = \pm \sigma(z,\bar z) = - \frac{g^2}{4\pi}  \sum_i\,q_i\, {\rm log} |z-z_i|^2~,
\end{equation}
while the solution in the magnetic theory is
\begin{equation}
\label{genconfig}
v_\pm = v_0^\pm \prod_i {\left( {z-z_i \over {\bar z - \bar z_i}}\right)^{\pm q_i/2}}~.
\end{equation}
A plot for a small 3$\times$3 periodic lattice in the magnetic theory appears in Figure \ref{fig:threebythree}.

In the IR electric theory, the solution (\ref{genconfig}) becomes
\begin{equation}
|\sigma|=2\pi |v_0^\pm|^2\,,\gamma= i \frac{g^2}{4\pi}\,\sum_i\,q_i \log \frac{z-z_i}{\bar z - \bar z_i}\,.
\end{equation}
There are basically several small cores around which there is a winding of the dual photon.

The ease with which one can find the analytical solutions for multiple-defect configurations, and analyze small fluctuations around the defect solutions, makes this a very promising system for investigating defect effects on linear response and transport phenomena in a strongly-coupled quantum gauge theory.

\section{Non-supersymmetric deformations and ground-state entropy}\label{sec:nonsusy}

It is easy to obtain controlled results for (slightly) non-supersymmetric theories by deforming any of the previous
results.  For instance, when a configuration is supersymmetric if $\rho(x) = \hat D(x)$ (sourcing $A_0$ and
$\sigma$ in the ${\cal N}=4$ QED theory), deforming slightly to 
\begin{equation}
\label{control}
\rho(x) = \hat D(x) + \delta \hat D,~\delta \hat D << \hat D
\end{equation}
allows a controlled expansion about the more precisely calculable supersymmetric results.

Here, we use this philosophy to provide a simple example of a vexing phenomenon which has shown up in the
study of finite density systems in AdS/CFT.  In that context, to turn on a finite density of a global charge in the
conformal field theory, one is instructed to study a charged black brane geometry in the bulk AdS space-time.
The simplest such systems -- charged black branes which arise in Einstein-Maxwell theory -- give rise to a 
puzzle.  The extremal black brane (which corresponds to the ground state of the doped
field theory) has a non-vanishing entropy at zero temperature.  The geometry of its near-horizon region is $AdS_2 \times \mathbb{R}^2$, and the fact that the horizon
is extended implies an (extensive) ground-state degeneracy.

This seems like a surprising result, as a doped, strongly interacting quantum field theory would not in general
be expected to have ground state entropy.  (For instance, this entropy violates `Nernst's theorem').  Discussions of this phenomenon
can be found in \cite{Kraus,Polchinski} and references therein.  Here, we point out that our simple mirror pairs
provide an example of this phenomenon where the field theories involved are explicitly known (and extremely
simple).  

Consider the $\mc N=4$ SQED theory in the presence of a constant density $\rho$ of external electric charges. This can be viewed as a limit of the case discussed in \S \ref{subsec:electricN4} where we have a lattice of electric charges and the theory is being analyzed at distances much larger than the lattice spacing. The classical electric potential grows with distance, $A_0 = \pm \sigma= \frac{\rho g^2}{8\pi}r^2$. The long distance physics is described by the mirror dual in in \S \ref{subsec:global-vorticesN4} with a constant magnetic field and D-term for the global $U(1)_J$. This is a supersymmetric version of the Landau level problem, where, because of the cancellation between the magnetic field and D-term sources, both the scalars and the fermions have Landau level wavefunctions with degenerate masses.\footnote{This system was also studied by~\cite{Polchinski}.}
The lowest Landau level has vanishing energy and preserves half of the supersymmetries. Recalling the discussion around (\ref{topcase}), the bosonic and fermionic wavefunctions satisfy the first order equations $D_{\bar z} v_\pm=0$ and $D_{\bar z} \psi_\pm=0$, or their complex conjugates depending on the preserved supersymmetry.


Instead, we choose
to break supersymmetry by turning on the $\hat B$-field but leaving $\hat D = 0$.
The condition for mirror symmetry to still be an approximate symmetry is that $\hat B$ be small in units of the gauge coupling,
an analogue of the condition (\ref{control}) that prevents the breaking of supersymmetry from leading to very large corrections to our statements.
Now in the theory with $\hat D = 0$ but $\hat B \neq 0$, the background magnetic field gaps the scalars in the free
hypermultiplet: they
live in Landau levels, but with a positive zero-point energy for the lowest Landau level (arising from the appropriate
harmonic oscillator wavefunction).

The fermions, on the other hand, still have gapless modes.  Their Hamiltonian is
\begin{equation}
H = \bar\psi (\gamma^z D_z + \gamma^{\bar z} D_{\bar z}) \psi~.
\end{equation}
The Hamiltonian vanishes for $\gamma^{\bar z} \psi = 0$ and $D_z \psi = 0$,
which comprise the lowest Landau level for the fermion.
An electron of charge $e$ in a magnetic field $\hat B$ has an orbit of size
\begin{equation}
\pi r^2 \sim {1\over e \hat B}
\end{equation}
which arises just by thinking about orbits of charged particles in a magnetic field.  This means that in a plane
of area ${\cal A}$, one can fit of order ${\cal A}\hat B$ electrons in the lowest Landau level (for a nice discussion in
the context of the quantum Hall effect, see \cite{Zee}).  

We conclude that a system can develop a ground state degeneracy due to strong dynamics, not seen in terms of the UV description. 
In a $2+1$--dimensional theory, all the spatial directions are threaded by a magnetic field, and the density of fermion zero modes gives rise to a ground state entropy, precisely along the lines envisioned in \cite{Kraus,Polchinski}.  The merit of
this example is that the strongly coupled system is a well-known field theory -- the supersymmetric analogue of 
2+1 dimensional quantum electrodynamics.\footnote{This system presents a counterexample to the well known Inverse Ninja Rule, in that the small $N$ gauge theory is more tractable than its large $N$ counterparts.  We thank E. Lim for discussions of this point.}

\section{D-brane picture}\label{sec:Dbrane}

There are simple D-brane constructions of the various 3d supersymmetric field theories we've studied \cite{HananyWitten,deBoer:1996mp,deBoer:1996ck}, including
the doping with external charges. This section reviews the D-brane realization of the $\mc N=4$ and $\mc N=2$ theories and realizes the SUSY defects in terms of semi-infinite F1 strings and D1 branes. This gives a geometric way of understanding some of previous discussion.

\subsection{Engineering the field theories}

The 3d ${\cal N}=4$ QED theory with a single flavor can be 
constructed by the type IIB string theory brane configuration depicted in Figure \ref{fig:electric1}.

\begin{figure}[h!]
\begin{center}
\begin{tabular}{c|ccc|ccc|c|ccc}
&$0$&$1$&$2$&$3$&$4$&$5$&$6$&$7$&$8$&$9$\\
\hline
D3  & x& x & x & & & & x & & & \\
2\,NS5& x& x & x & x& x& x&  & & & \\
D5& x& x & x&  & & & & x& x& x \\
\end{tabular}
\end{center}
\caption{\small{The brane configuration engineering the 3d ${\cal N}=4$ QED theory on a stack of D3-branes (stretching between NS5 branes in the $x_6$ direction).}}
\label{fig:electric1}
\end{figure}

The D5-brane location coincides with the D3-branes in the 345 directions, with the D3-D5 strings giving rise to the massless hypermultiplet ${\cal Q}$.  The Coulomb branch of vacua is parametrized by the D3 location along the NS5s
in the 345 directions; this geometrizes $\sigma$ and $\Phi$, but the dual photon is not geometrized by the brane construction. 

The mirror brane configuration is obtained by the  S-duality transformation of type IIB string theory.  This maps D3 branes to themselves, but takes NS5 branes to D5 branes and vice-versa, as in Figure \ref{fig:magnetic1}.

\begin{figure}[h!]
\begin{center}
\begin{tabular}{c|ccc|ccc|c|ccc}
&$0$&$1$&$2$&$3$&$4$&$5$&$6$&$7$&$8$&$9$\\
\hline
D3  & x& x & x & & & & x & & & \\
2\,D5& x& x & x & x& x& x&  & & & \\
NS5& x& x & x&  & & & & x& x& x \\
\end{tabular}
\end{center}
\caption{\small{The S-dual brane configuration to Figure \ref{fig:electric1}, giving the theory of a free hypermultiplet.}}
\label{fig:magnetic1}
\end{figure}

As the boundary conditions freeze the D3 location in the 345 directions when it is suspended between D5s, this gives rise to a field theory with no Coulomb branch.  By SUSY, one can infer (or compute directly) that there are no ${\cal N}=4$ vector multiplets in this dual theory.  However, the `half' D3s can split their locations along the NS5 brane.  This gives rise to a Higgs branch of vacua, as the halves split to different locations in the 345 directions.  (Again, one direction in the moduli space of vacua is not geometrized).

The 3d ${\cal N}=2$ configurations are obtained in a similar manner.  Rotating one of the NS5 branes into an NS5' brane (which wraps the 89 directions instead of the 45 directions) breaks the SUSY in a suitable way.  The Coulomb branch is partially lifted.  The D3 may no longer slide in the 45 directions while still ending on the NS5 and NS5' branes.  However, the $\sigma$ modulus still exists (from translations of the D3 along the 3 direction).  In addition, there is a new Higgs branch of vacua. Geometrically, this corresponds to moving a D3 segment between the D5 and NS5' along the 89 directions.
In the magnetic dual, one of the D5s in Figure \ref{fig:magnetic1} is rotated into a D5' along the 89.

\subsection{Including the background sources}

The basic reason that we can easily describe our backgrounds in terms of brane sources is that the delta function sources (which can give arbitrary sources by use of the appropriate Green's function) are geometrized nicely by string theory.
For instance, in the ${\cal N}=4$ QED theory, the basic charge which sources $A_0$ and $\sigma$ is in fact the fundamental string ending on the D3-brane, as in Figure \ref{fig:electricF1} below.  This result is of course well known in the literature
on Wilson loops in string theory. 

\begin{figure}[h!]
\begin{center}
\begin{tabular}{c|ccc|ccc|c|ccc}
&$0$&$1$&$2$&$3$&$4$&$5$&$6$&$7$&$8$&$9$\\
\hline
D3  & x& x & x & & & & x & & & \\
2\,NS5& x& x & x & x& x& x&  & & & \\
D5& x& x & x&  & & & & x& x& x \\
\hline
F1  & x& &  &x & & &  & & &
\end{tabular}
\end{center}
\caption{\small{The brane configuration for a delta function electric source in ${\cal N}=4$ QED, as described in \S \ref{subsec:electricN4}.}}
\label{fig:electricF1}
\end{figure}

The fact that the sources couple to $A_0$ and $\sigma$ is here a simple consequence of the F1 worldvolume being extended along the $0$ and $3$ directions (the later corresponds to the $\sigma$ field in the $SO(3)$ parametrization we have chosen).
It is then automatic to find the magnetic dual - one simply applies type IIB S-duality to the brane configuration including the source.  The result is in Figure \ref{fig:magneticD1}.

\begin{figure}[h!]
\begin{center}
\begin{tabular}{c|ccc|ccc|c|ccc}
&$0$&$1$&$2$&$3$&$4$&$5$&$6$&$7$&$8$&$9$\\
\hline
D3  & x& x & x & & & & x & & & \\
2\,D5& x& x & x & x& x& x&  & & & \\
NS5& x& x & x&  & & & & x& x& x \\
\hline
D1  & x& &  &x & & &  & & &
\end{tabular}
\end{center}
\caption{\small{The mirror `magnetic' configuration, an external vortex in the theory of a free hypermultiplet. }}
\label{fig:magneticD1}
\end{figure}

Similar IIB brane configurations geometrizing the external magnetic defect in ${\cal N}=4$ QED, and its mirror electric source for $U(1)_J$, are shown in Figures \ref{fig:electricD1} and \ref{fig:magneticF1}. The insertion of the semi-infinite D1 brane in the electric theory realizes the localized FI term introduced in \S \ref{subsec:magneticN4}. In the D-brane picture, the D1 brane pulls on one of the NS5 branes along along its worldvolume direction 7, which is an FI term localized around the point where the brane is inserted. (The same effect along the time direction gives rise to a localized external $\hat A_0$). As a limit of this, a smeared density of D1 branes would give rise to a uniform displacement of  an NS5 brane, which is the usual constant FI term.

\begin{figure}[h!]
\begin{center}
\begin{tabular}{c|ccc|ccc|c|ccc}
&$0$&$1$&$2$&$3$&$4$&$5$&$6$&$7$&$8$&$9$\\
\hline
D3  & x& x & x & & & & x & & & \\
2\,NS5& x& x & x & x& x& x&  & & & \\
D5& x& x & x&  & & & & x& x& x \\
\hline
D1  & x& &  & & & &  &x & &
\end{tabular}
\end{center}
\caption{\small{The brane configuration for a (localized) source of $B$ and a (localized) FI term in ${\cal N}=4$ QED.} }
\label{fig:electricD1}
\end{figure}

\begin{figure}[h!]
\begin{center}
\begin{tabular}{c|ccc|ccc|c|ccc}
&$0$&$1$&$2$&$3$&$4$&$5$&$6$&$7$&$8$&$9$\\
\hline
D3  & x& x & x & & & & x & & & \\
2\,D5& x& x & x & x& x& x&  & & & \\
NS5& x& x & x&  & & & & x& x& x \\
\hline
F1  & x& &  & & & &  &x & &
\end{tabular}
\end{center}
\caption{\small{The source for an `electric' charge of $U(1)_J$ in the mirror theory.}}
\label{fig:magneticF1}
\end{figure}

In all cases, the semi-infinite F1s and D1s behave as external sources in the field theory because of their infinite mass.
Solving for the field configurations generated by these sources (by finding bulk supergravity solutions incorporating `brane bending') should reproduce the direct field theory considerations of \S \ref{sec:defectsN4}.  The ${\cal N}=2$ theories with external sources can be engineered in the obvious similar manner, and we do not discuss them here.

\section{Discussion and future directions}\label{sec:discussion}

In this paper, we have started to explore the fact that defects in supersymmetric theories can preserve half of the supersymmetry, giving rise to models of defects interacting with a strongly coupled gauge theory that can be surprisingly tractable. This includes supersymmetric models at finite density or in the presence of magnetic flux. There are many directions for further exploration.

A crucial role in our analysis was played by the mirror symmetry of 3d supersymmetric gauge theories,
which in the particular case of 3d ${\cal N}=4$ QED maps questions about defect interactions with the
strongly coupled IR theory to dual questions in a free magnetic description. This allows one to write down simple explicit solutions reflecting the backreaction of arbitrary arrays of defect charges
on the theory, as in \S \ref{sec:lessons}. Further exploration of these configurations, in particular of linear response
and transport in the presence of a defect lattice, could prove interesting.

Our analysis also suggested possible generalizations of the Abrikosov-Nielsen-Olesen vortices to a spatially varying FI term and magnetic flux source, which would be very interesting to analyze.
We discussed only the simplest mirror pairs of theories with 3d ${\cal N}=4$ and ${\cal N}=2$ supersymmetry. Much richer collections of mirror pairs are known~\cite{Mirror, deBoer:1996mp,deBoer:1996ck,deBoer:1997kr,Aharony:1997bx,Kapustin:1999ha},
and new phenomena may be visible there.  Similarly, systems with Chern-Simons terms are
of considerable interest in condensed matter physics, and the exploration of analogous
defect configurations to the ones we discussed here
in supersymmetric Chern-Simons theories (for relatively comprehensive recent discussions, see \cite{Ken}) should be
straightforward.

We focused here on theories in 2+1 space-time dimensions. But the existence of supersymmetry preserving defects is common to theories with BPS particles, including 4d ${\cal N}=2$ theories and
various 2d supersymmetric theories. The key is that there should be a scalar in the gauge multiplet that can cancel the SUSY-variation due to the finite charge density. Exploring such defect models,
perhaps using the tools of duality in other dimensions, is likely to be worthwhile.

\section*{Acknowledgments}

We would like to thank K.~Intriligator, S.~Sachdev and D.~Tong for useful comments on a draft of this work, and T.~Cohen for collaboration on an early attempt at this analysis.
The research of S.K. and G.T.~is supported in part by the National Science Foundation under grant no.~PHY-0756174.  S.K. is also supported by the Department of Energy under contract
DE-AC02-76SF00515, and the John Templeton Foundation.
 A.H.~is supported by the Department of Energy under contract DE-SC0009988.

\end{document}